\documentclass[12pt]{article}
\usepackage{a4wide,epsfig}
\usepackage{graphicx}

\usepackage{amsmath} 

\newcommand{\alphaMOM}{\alpha^{\rm MOM}}
\newcommand{\betaMOM}{\beta^{\rm MOM}}
\newcommand{\PiMOM}{\Pi^{\rm MOM}}

\newcommand{\lnfourZ}{\ln(4Z)}
\newcommand{\lnfourZtwo}{\ln^2(4Z)}

\def\bbuildrel#1_#2^#3%
{\mathrel{\mathop{\kern 0pt#1}\limits_{#2}^{#3}}}

\newcommand{\ice}[1]{\relax}

\newcommand{\beq}{\begin{equation}}
\newcommand{\eeq}{\end{equation}}
\newcommand{\bea}{\begin{eqnarray}}
\newcommand{\eea}{\end{eqnarray}}

\newcommand{\ba}{\begin{array}} 
\newcommand{\ea}{\end{array}}

\newcommand{\alsVI}{\alpha_s^{(6)}}

\newcommand{\alspmu}{\frac{\alpha_s^{(n_f)}(\mu^2)}{\pi}}

\newcommand{\AlspmuVI}{\frac{\alpha^{(6)}_s(\mu^2)}{4 \pi}}

\newcommand{\alsMOM}{\alpha_s^{\mathrm{MOM}} }

\newcommand{\alspMOM}{\frac{\alpha_s^{\mathrm{MOM}}}{\pi} }

\newcommand{\alsMOMbar}{\alpha_s^{\overline{\mathrm{MOM}}} }

\newcommand{\ovl}{\overline}

\newcommand{\msbar}{\overline{\mbox{MS}}}

\begin{document}    

%- {{{ Title, abstract

\title{\vskip-3cm{\baselineskip14pt
\centerline{\normalsize\hfill DESY 08--184}
\centerline{\normalsize\hfill SFB/CPP-08-92}
\centerline{\normalsize\hfill TTP08--50}
}
\vskip.7cm
Strong-Coupling Constant at Three Loops
in Momentum Subtraction Scheme
}

\author{
{K.G. Chetyrkin}$^{a,b}$,
{B.A. Kniehl}$^{c}$
and
{M. Steinhauser}$^a$
  \\[1em]
  {\normalsize (a) Institut f\"ur Theoretische Teilchenphysik,}\\
  {\normalsize Universit\"at Karlsruhe (TH),}
  {\normalsize Karlsruhe Institute of Technology (KIT),}\\
  {\normalsize 76128 Karlsruhe, Germany}
 \\[.5em]
{\normalsize (b) Institute for Nuclear Research,
 Russian Academy of Sciences,}\\ 
{\normalsize Moscow 117312, Russia}
 \\[.5em]
  {\normalsize (c) II. Institut f\"ur Theoretische Physik,}\\ 
  {\normalsize Universit\"at Hamburg, 22761 Hamburg, Germany}
}
 
\date{}
\maketitle

\begin{abstract}
\noindent
In this paper we compute the three-loop corrections to the $\beta$
function in a momentum subtraction (MOM) scheme with a massive quark.
The calculation is performed in the background field formalism
applying asymptotic expansions for small and large momenta.
Special emphasis is devoted to the relation between the 
coupling constant in the MOM and $\overline{\rm{MS}}$ schemes
as well as their ability to describe the phenomenon of decoupling.

It is demonstrated by an explicit comparison that the
$\overline{\rm{MS}}$ scheme can be consistently used to relate the
values of the MOM-scheme strong-coupling constant in the energy
regions  higher and lower than the massive-quark production
threshold. This procedure obviates  the necessity to know the full
mass dependence of the MOM $\beta$ function and clearly demonstrates
the equivalence of both schemes for the description of physics outside
the threshold region.

\medskip


\end{abstract}

\thispagestyle{empty}
\newpage
\setcounter{page}{1}

\renewcommand{\thefootnote}{\arabic{footnote}}
\setcounter{footnote}{0}

%- }}}
%- {{{ Introduction

\section{Introduction}

Within the perturbative framework, the $\overline{\rm MS}$
scheme~\cite{tHooft:1973mm,Bardeen:1978yd} based on dimensional
regularization~\cite{tHooft:1972fi,Ashmore:1972uj,Cicuta:1972jf} is a
well-established scheme for the renormalization of fields and
parameters. This applies in particular to $\alpha_s$,
the coupling constant of Quantum Chromodynamics (QCD).
One of the major advantages of the $\overline{\rm MS}$ scheme is its
simplicity in practical applications. The main reason for this is
that it belongs to the class of so-called mass-independent schemes
where the renormalization constants are independent of the precise
configuration of masses and external momenta involved in the problem.

Within the $\overline{\rm MS}$ scheme, the beta function 
governing the running of $\alpha_s$
is know in the four-loop
approximation~\cite{Vermaseren:1997fq,Czakon:2004bu}. 
In order to correctly account for the heavy-quark thresholds,
also the corresponding matching (or decoupling) conditions
are needed, which allows for a precise relation of $\alpha_s$ at
widely separated energy scales like, e.g., the tau lepton and $Z$
boson masses. 
Four-loop running goes along with three-loop matching, which is also
known since more than ten
years~\cite{Chetyrkin:1997un}.\footnote{Recently, also the 
  four-loop decoupling constants have been
  computed~\cite{Schroder:2005hy,Chetyrkin:2005ia}.}

Other renormalization schemes which do not have the nice property of
mass-independence are significantly more complicated from the 
technical point of view --- mainly because one has to deal with
Feynman integrals involving many mass scales.
Still, at the level of precision which has been reached in the recent
years, it is necessary to have a cross check of the dependence on the
renormalization scheme. In this paper, we want to provide an
alternative set-up to the running and decoupling of $\alpha_s$ in the 
$\overline{\rm MS}$ scheme and consider a momentum subtraction (MOM)
scheme for the definition of $\alpha_s$. We will provide 
$\overline{\rm MS}$ to MOM conversion
formulae and the MOM beta function in the
three-loop order and are thus able to cross check the 
$\overline{\rm MS}$ running of $\alpha_s$.
A two-loop analysis has been performed in
Ref.~\cite{Jegerlehner:1998zg}. In this paper we check the calculation
of Ref.~\cite{Jegerlehner:1998zg} and extend the analysis to three loops.

The remainder of the paper 
is organized as follows: In the next section, we
describe our setup. In particular, we derive the relation between the strong
coupling in the $\overline{\rm MS}$ scheme and in two versions of the momentum
subtraction scheme and provide the corresponding beta functions.
In Section~\ref{sec::results}, we present our analytical results for the gluon
polarization function in the background field formalism and discuss the
phenomenological applications in Section~\ref{sec::applications}, where we
compare the running in the $\overline{\rm MS}$ and MOM schemes.
Our conclusions are summarized in Section~\ref{sec::concl}.

%- }}}
%- {{{ Coupling in the MOM scheme

\section{The strong coupling in the MOM scheme}

For convenience, we adopt Landau gauge, which has the advantage that
the renormalization group equations for the gauge parameter and
$\alpha_s$ decouple. Furthermore, we require that the 
polarization function of the gluon vanishes for $Q^2\equiv-q^2=\mu^2>0$.

For the practical calculation, we adopt the background field
gauge~\cite{Abbott:1980hw}, which has the 
nice feature that the $\beta$ function of the strong coupling is 
determined from the gluon polarization function alone.
The latter is given by
\begin{eqnarray}
  \Pi^{\mu\nu}(q) &=& \left(-g^{\mu\nu}q^2 + q^\mu q^\nu\right)\Pi(q^2)
  \,,
  \label{eq::pimunu}
\end{eqnarray}
which is conveniently decomposed as follows
\begin{eqnarray}
  \Pi(q^2) &=& \sum_{i\geq1} \Pi^{(i)}(q^2,\mu^2,\{M_Q^2\})
  \left(\frac{\alpha_s}{\pi}\right)^i
  \,.
  \label{eq::pi}
\end{eqnarray}
In the $i$-loop contribution, the dependence on $q$,
$\mu$ and the various quark masses is explicitly displayed.
Formulae~(\ref{eq::pimunu}) and~(\ref{eq::pi})
hold both in the $\overline{\rm MS}$ and MOM
schemes. The corresponding functions, $\Pi(q^2)$ and 
$\PiMOM(q^2)$, can be used to obtain
a relation between $\alpha_s$ and $\alphaMOM_s$, the strong couplings
in the $\overline{\rm MS}$ and MOM schemes, using
the fundamental concept of the {\em invariant charge }
 \cite{Bogolyubov:1956gh,Shirkov:1998ak}:
\begin{eqnarray}
  \frac{\alphaMOM_s(\mu^2)}{1+\PiMOM(q^2)}
  =
  \frac{\alpha_s(\mu^2)}{1+\Pi(q^2)}
  \label{eq::PiMOMvsMS}
  \,.
\end{eqnarray}
It is an important and unique  feature of the background field gauge that
the invariant charge  is expressible in terms of 
the coupling constant and the gluon polarization  operator {\em only}
in exactly the same  simple way as in QED.
We define $\PiMOM(q^2)$ such that $\PiMOM(-\mu^2)=0$ and,
consequently, we have
\begin{eqnarray}
 \alphaMOM_s(\mu^2) &=& \alpha_s^{(n_f)}(\mu^2)
    \left[1
  +
  c_1 \alspmu + c_2 \left( \alspmu \right)^2 
  +
  c_3 \left( \alspmu \right)^3 \right]
  \,,
  \nonumber\\
  c_1 &=& -\Pi_0^{(1)},
  \nonumber\\
  c_2 &=& -\Pi_0^{(2)} + \left(\Pi_0^{(1)}\right)^2,
  \nonumber\\
  c_3 &=& -\Pi_0^{(3)} + 2\Pi_0^{(1)}\Pi_0^{(2)}
      - \left(\Pi_0^{(1)}\right)^3 
  \label{eq::asMOMvsMS}
  \,,
\end{eqnarray}
where $\Pi_0^{(i)}=\Pi^{(i)}(-\mu^2)$ has been introduced.
It is instructive to look at the explicit expressions in the massless
limit with $n_f=n_l$ massless quarks. In this case, we obtain
\begin{eqnarray}
  \alphaMOM_s(\mu^2) &=& \alpha_s^{(n_l)}(\mu^2)
    \Bigg[
      1
      + \frac{\alpha_s^{(n_l)}(\mu^2)}{4\pi} 
      \left(
      \frac{205}{12} - \frac{10}{9}n_l
      \right)
      + \left(\frac{\alpha_s^{(n_l)}(\mu^2)}{4\pi}\right)^2 
      \left(
      \frac{90391}{144}
      \right.\nonumber\\&&\left.\mbox{}
      - \frac{513}{8}\zeta(3)
      + \left(-\frac{2066}{27} - \frac{4}{3}\zeta(3)\right) n_l
      + \frac{100}{81} n_l^2
      \right)
      \nonumber\\&&\mbox{}
      + \left(\frac{\alpha_s^{(n_l)}(\mu^2)}{4\pi}\right)^3 
      \left(
      \frac{50765707}{1728} 
      - \frac{23343}{4}\zeta(3)
      - \frac{24885}{64}\zeta(5)
      + \left(-\frac{860917}{162} 
      \right.\right.\nonumber\\&&\left.\left.\mbox{}
      + \frac{20423}{54}\zeta(3)
      + \frac{2320}{9}\zeta(5)
      \right)n_l
%      \right.\nonumber\\&&\left.\mbox{}
      + \left(\frac{209407}{972} + \frac{28}{9}\zeta(3)\right) n_l^2
      - \frac{1000}{729} n_l^3
      \right)
      \Bigg]\,.
    \nonumber\\
    \label{eq::asMOM2asMS}
\end{eqnarray}

In analogy to the $\overline{\rm MS}$ scheme,
the $\beta$ function in the MOM scheme is defined through
\begin{eqnarray}
  \mu^2\frac{{\rm d}}{{\rm d}\mu^2}
  \frac{\alphaMOM_s}{\pi}
  &=& \betaMOM(\alphaMOM_s)
  \,\,=\,\, - \left(\frac{\alphaMOM_s}{\pi}\right)^2
  \sum_{i\geq0}
  \betaMOM_i\left(\frac{\alphaMOM_s}{\pi}\right)^i
  \,,
  \label{eq::betadef}
\end{eqnarray}
where --- in contrast to the $\overline{\rm MS}$ scheme ---
the coefficients $\betaMOM_i$ are functions of 
the renormalization scale $\mu$ and the quark masses.
With the help of Eq.~(\ref{eq::PiMOMvsMS}), where we replace on the
right-hand side the $\overline{\rm MS}$ renormalized quantities by the bare
ones, it is possible to obtain a relation between $\betaMOM$ and the
coefficients $\Pi^{(i),{\rm MOM}}=\Pi^{(i),{\rm MOM}}(q^2,\mu^2,\{M_Q^2\})$,
which reads
\begin{eqnarray}
  \betaMOM(\alphaMOM_s) &=&
  \frac{\alphaMOM_s}{\pi}
  \frac{ 
    \sum_{i\ge1} \left( \frac{\alphaMOM_s}{\pi} \right)^i 
    \mu^2\frac{{\rm d}}{{\rm d}\mu^2} \Pi^{(i),{\rm MOM}}
  }
  { 1 - \sum_{i\ge1} (i-1) \left( \frac{\alphaMOM_s}{\pi} \right)^i 
    \Pi^{(i),{\rm MOM}} 
  }
  \,.
  \label{eq::betaMOM}
\end{eqnarray}
From this equation, one can easily derive convenient formulae for
$\betaMOM_i$.
Note that the term in the denominator of Eq.~(\ref{eq::betaMOM}) 
contributes  for the first time at the three-loop order. Let us also mention
that,
starting at this order, a non-trivial $q^2$ dependence occurs on the
right-hand side of Eq.~(\ref{eq::betaMOM}) which has to cancel in 
the proper combination of the $\Pi^{(i),{\rm MOM}}$ functions.

The functions $\betaMOM_0$ and $\betaMOM_1$ are known
analytically~\cite{Jegerlehner:1998zg}.
The three-loop contribution $\betaMOM_2$ is evaluated in the asymptotic
regions for large and small quark masses analytically in this paper.
An approximate formula valid for arbitrary quark masses is easily
obtained by interpolation between the low- and high-energy regions.

In the massless limit, the first three coefficients are given by
\begin{eqnarray}
  \betaMOM_{0,\rm ml} &=&
  \frac{1}{4} \left[
    \frac{11}{3}C_A - \frac{4}{3}Tn_l
  \right]
  \,,
  \nonumber\\
  \betaMOM_{1,\rm ml} &=&
  \frac{1}{16} \left[
    \frac{34}{3}C_A^2
    - \frac{20}{3} C_ATn_l
    - 4 C_F T n_l
  \right]
  \,,
  \nonumber\\
  \betaMOM_{2,\rm ml} &=&
  \frac{1}{64} \left[
    - \left( - \frac{3005}{24} + \frac{209}{8} \zeta(3) \right) C_A^3
    - \left( \frac{1861}{18} + \frac{119}{6} \zeta(3) \right) C_A^2 T n_l
    \right.\nonumber\\&&\left.\mbox{}
    - \left( \frac{605}{9}   - \frac{176}{3} \zeta(3) \right) C_A C_F T n_l
    - \left(- \frac{130}{9}  - \frac{32}{3} \zeta(3) \right) C_A T^2 n_l^2
    \right.\nonumber\\&&\left.\mbox{}
    - \left(- \frac{184}{9}  + \frac{64}{3} \zeta(3) \right) C_F T^2 n_l^2
    + 2 C_F^2 Tn_l
  \right]
  \,,
  \label{eq::beta012}
\end{eqnarray}
where $C_A=3, C_F=4/3, T=1/2$ and $n_l$ is the number of massless quarks.
Since the first two coefficients of the $\beta$ function are scheme
independent $\betaMOM_{0,\rm ml}$ and $\betaMOM_{1,\rm ml}$ coincide with
their counterparts in the $\overline{\rm MS}$ scheme.
$\betaMOM_{2,\rm ml}$, however, differs from its $\overline{\rm MS}$
counterpart~\cite{Tarasov:1980au,Larin:1993tp}. 
It is worthwhile to mention that $\betaMOM_{2,\rm ml}$ contains 
the Riemann $\zeta$ function $\zeta(3)$, which in the 
$\overline{\rm MS}$ scheme only appears at the four-loop order.

The three-loop results in Eqs.~(\ref{eq::asMOM2asMS})
and~(\ref{eq::beta012}) are new, and the two-loop expressions are in
agreement with Ref.~\cite{Jegerlehner:1998zg}.

The practical evaluation of $\PiMOM(q^2)$ entering the equation for the beta
function can be reduced to the evaluation of $\Pi(q^2)$ in the 
$\overline{\rm MS}$ scheme. The corresponding relation is obtained from 
Eq.~(\ref{eq::PiMOMvsMS}), this time for arbitrary values of
$q^2$ and $\mu^2$,
which can be solved for $\PiMOM$. After properly replacing
$\alphaMOM_s$ by $\alpha_s$ using Eq.~(\ref{eq::asMOMvsMS}),
one gets (the dependence on the quark masses is suppressed)
\begin{eqnarray}
  \Pi^{(1),{\rm MOM}}(q^2) &=& \Pi^{(1)}(q^2) - \Pi_0^{(1)}
  \,,
  \nonumber\\
  \Pi^{(2),{\rm MOM}}(q^2) &=& \Pi^{(2)}(q^2) - \Pi_0^{(2)}
  \,,
  \nonumber\\
  \Pi^{(3),{\rm MOM}}(q^2) &=& \Pi^{(3)}(q^2) - \Pi_0^{(3)} 
                     + \Pi_0^{(1)} \left( \Pi^{(2)}(q^2) - \Pi_0^{(2)} \right)
  \,,
\end{eqnarray}
where $\Pi_0^{(i)}$ is defined below Eq.~(\ref{eq::asMOMvsMS}).
Note, that by construction we have $\Pi^{\rm MOM}(-\mu^2)=0$.

The polarization function in the $\overline{\rm MS}$ scheme
is obtained in the standard way by renormalizing $\alpha_s$ in the
$\overline{\rm MS}$ scheme, the quark masses in the on-shell
scheme and taking care of the gluon wave function renormalization.

In Ref.~\cite{Jegerlehner:1998zg}, it has been observed that 
there are relatively large coefficients in the
relation between $\alpha_s$ and $\alpha_s^{\mathrm{MOM}}$
when running from $\alpha_s(M_Z)$ down to, say,
$\alpha_s(M_{\tau})$.
The situation was improved in Ref.~\cite{Jegerlehner:1998zg} by a simple trick
of rescaling the scale parameter $\mu$.  Let us start from the massless limit
corresponding to $\mu \gg M_t$. In this case, relation (\ref{eq::asMOM2asMS})
assumes the form
\begin{eqnarray}
  \alsMOM(\mu^2) &=& \alpha_s^{(6)}(\mu^2)\Bigg[
  1 
  + 10.417\,\frac{\alpha_s^{(6)}(\mu^2)}{4\pi} 
  + 126.350\,\left(\frac{\alpha_s^{(6)}(\mu^2)}{4\pi}\right)^2
  \nonumber\\&&\mbox{}
  + 2000.062\,\left(\frac{\alpha_s^{(6)}(\mu^2)}{4\pi}\right)^3
  \Bigg]
  \,.
  \label{asMOM_through_asMS:nl6,nl}
\end{eqnarray}
In a next step, we introduce a new, rescaled MOM scheme with the help of
\begin{eqnarray}
  \alsMOMbar(\mu^2) &\equiv&  \alsMOM(x_0^2\mu^2)
  \label{rescale:def1}
\end{eqnarray}
or, equivalently (with $L = \ln(x_0^2)$),
\begin{eqnarray}
  \alsMOMbar &\equiv& \alsMOM \left[ 1
  +
  r_1 \alspMOM + r_2 \left( \alspMOM \right)^2 
  +
  r_3 \left( \alspMOM \right)^3 \right]
  \,,
  \nonumber\\
  r_1 &=& -L\betaMOM_{0, \rm ml}\,,
  \nonumber\\
  r_2 &=& \left(L \betaMOM_{0, \rm ml}\right)^2 -  L \betaMOM_{1, \rm ml}
  \,,
  \nonumber\\
  r_3 &=& \frac{5}{2} L^2 \betaMOM_{0,\rm ml} 
  \betaMOM_{1,\rm ml} -  L \betaMOM_{2,\rm ml} - \left(L \betaMOM_{0,\rm
      ml}\right)^3 
  \label{rescale:def2}
  \,,
\end{eqnarray}
where $\betaMOM_{i,\rm ml}$ are given in Eq.~(\ref{eq::beta012}).
The corresponding generalization of Eq.~(\ref{eq::asMOMvsMS}) reads:
\begin{eqnarray}
\alsMOMbar(\mu^2) &=& \alpha_s^{(n_f)}(\mu^2)
    \left[1
  +
  \overline{c}_1 \alspmu + \overline{c}_2 \left( \alspmu \right)^2 
  +
  \overline{c}_3 \left( \alspmu  \right)^3 \right]
  \,,
  \nonumber\\
  \overline{c}_1 &=& r_1 -\Pi_0^{(1)},
  \nonumber\\
  \overline{c}_2 &=&  -\Pi_0^{(2)} + \left(\Pi_0^{(1)}\right)^2  -2\,\Pi_0^{(1)}\,r_1   + r_2,
  \nonumber\\
  \overline{c}_3 &=&  -\Pi_0^{(3)} 
  + 2\Pi_0^{(1)}\Pi_0^{(2)}
  - \left(\Pi_0^{(1)}\right)^3 
  + 3 \left(\Pi_0^{(1)} \right)^2 r_1
  \nonumber\\&&\mbox{}  - 2 \Pi_0^{(2)} r_1
  -3\, \, \Pi_0^{(1)}\,r_2
  +r_3
  \label{eq::asMOMBvsMS}
  \,.
\end{eqnarray}
\ice{
amb2as = amb -> as + as^2*(c1 + r1) + as^3*(c2 + 2*c1*r1 + r2) +
      as^4*(c3 + c1^2*r1 + 2*c2*r1 + 3*c1*r2 + r3)
}
In a next step, following Ref.~\cite{Jegerlehner:1998zg}, we {\em tune} the
parameter $x_0$ so that the difference between $\alsMOMbar$ and $\alsVI$
starts only in order $\alpha_s^2$. The result reads\footnote{Note that there
  seems to be a misprint in the numerical value of $x_0$ quoted in
  Ref.~\cite{Jegerlehner:1998zg},
  however, in the caption of Fig.~4 therein it is correct.}
\begin{eqnarray}
  \ln(x_0^2) &=& \frac{125}{84}\,, \qquad  x_0  \approx 2.1044\,,
\end{eqnarray}
which leads to
\begin{eqnarray}
  r_1 &\approx& -2.60417   \,,\nonumber\\
  r_2 &\approx& 4.3635  \,,\nonumber\\
  r_3 &\approx& 2.2313  \,.
\end{eqnarray}

It is instructive to look again at the relation between $\alsMOMbar$ and
$\alsVI$ for $\mu\gg M_t$, which is now given by
\begin{eqnarray}
  \alsMOMbar(\mu^2) &=& \alsVI(\mu^2)\left\{
  1
  +  
  \tilde{k}_1 \AlspmuVI
  +
  \tilde{k}_2 \left(\AlspmuVI\right)^2
  +
  \tilde{k}_3
  \left(\AlspmuVI\right)^3
  \right\}
  \label{asMOM_through_asMS:nl6,nl,rescaled}
  \,,
\end{eqnarray}
where 
\begin{eqnarray}
  \tilde{k}_1 &=& 0,
  \nonumber\\ 
  \tilde{k}_2 &=& \frac{11063}{168} - \frac{577}{8}\zeta(3) \,\,\approx\,\,
  - 20.8472
  \,,
  \nonumber\\
  \tilde{k}_3  &=& 
  \frac{101389}{126} - \frac{345779}{288}\zeta(3) +
  \frac{222305}{192}\zeta(5) 
  \,\,\approx\,\,  562.0541
  \,.
\end{eqnarray}
As compared to Eq.~(\ref{asMOM_through_asMS:nl6,nl}), 
we observe a significant reduction in the magnitude of the
coefficients in the rescaled relation
(\ref{asMOM_through_asMS:nl6,nl,rescaled}), 
both at the two- and three-loop orders.\footnote{Note that our
  value for the two-loop coefficient in
  Eq.~(\ref{asMOM_through_asMS:nl6,nl,rescaled})
  ($-20.8472$) differs from the one obtained in
  Ref.~\cite{Jegerlehner:1998zg} ($-32.46$).}
Furthermore, there is a different sign in the three-loop coefficient
as compared to the two-loop one, which points to a better convergence of
the perturbative expansion.

In the massless limit, both definitions for $\alsMOMbar(\mu^2)$, namely
Eqs.~(\ref{rescale:def1}) and (\ref{rescale:def2}), are completely
equivalent. Following again Ref.~\cite{Jegerlehner:1998zg}, we choose
Eq.~(\ref{rescale:def2}) as the proper definition of the
$\alsMOMbar(\mu^2)$ for {\em all} values of $\mu$. 
This choice has the advantage that the 
thresholds in the corresponding function
$\beta^{\ovl{\mathrm{MOM}}}$ remain ``physical'',
that is located at $-\mu^2 = 4M_Q^2$. This 
follows directly from the relation between both $\beta$ functions:
\begin{eqnarray}
  \beta^{\ovl{\rm MOM}}_0 &=& \betaMOM_0 \,,
  \nonumber\\
  \beta^{\ovl{\rm MOM}}_1 &=& \betaMOM_1 \,,
  \nonumber\\
  \beta^{\ovl{\rm MOM}}_2 &=& \betaMOM_2 - r_1\betaMOM_1 
  +\left(r_2- r^2_1\right)\betaMOM_0 \,.
\end{eqnarray}

\ice{
                       2    2                       2            3
Out[7]= be1 h + b be2 h  + b  (be3 - be2 c1 - be1 c1  + be1 c2) h
 }

It is interesting to remark that the rescaling procedure significantly
improves the $\ovl{\rm MOM}$ to $\msbar$ relations also for moderate
and even rather low values of $\mu$ (see below). 

In the applications of Section~\ref{sec::applications}, we consider
the strong coupling both for energy scales of the order of or larger
than and for those significantly smaller than the top-quark mass.
In the latter case, we construct different MOM and 
$\overline{\rm MOM}$ schemes, which are derived from the choice $n_f=5$ 
as the massless limit. In this case, we obtain the values
\begin{eqnarray}
  x_0 &=& e^{415/552} \approx 2.1208\,,\nonumber\\
  \tilde{k}_2 &=& \frac{140689}{1656} - \frac{1699}{24}\zeta(3)
  \approx -0.1385\,,\nonumber\\
  \tilde{k}_3 &=& \frac{143409281}{89424} 
  - \frac{1225793}{864}\zeta(3) + \frac{518435}{576}\zeta(5) 
  \approx 831.5896 \,,\nonumber\\
  r_1 &=& -\frac{415}{144} \approx -2.8819\,,\nonumber\\
  r_2 &=& \frac{2228135}{476928} \approx 4.6719\,,\nonumber\\
  r_3 &=& -\frac{1188703175}{68677632} + \frac{705085}{55296}\zeta(3)
  \approx -1.9809\,. 
  \label{eq::nfMOM5values}
\end{eqnarray}
The same comments and conclusions hold as for $n_f=6$.

%- }}}
%- {{{ Results

\section{\label{sec::results}Results}

Let us in a first step briefly describe the evaluation of the gluon
polarization function up to three loops within the background field
formalism involving heavy quarks with generic mass $M_Q$.  The basic
idea is to evaluate $\Pi(q^2)$ for large and small external momenta
and to obtain an approximation for all values of $q^2/M_Q^2$ by a simple
interpolation procedure. Note that in our case the external momentum
is space-like so that there are no problems with particle thresholds.
Up to the two-loop order, only one quark flavour can occur in a
diagram. At three loops, there are diagrams with a second closed
fermion loop so that in principle a further mass scale can occur
(see, e.g., the diagram in Fig. 1(g)).
However, we assume a strong hierarchy in the quark masses such that we
can always neglect the lighter mass. Thus, in this section, we consider
QCD with total number $n_f$ of quark flavours.  One quark, $Q$, has
the (pole) mass $M_Q$, and all other $n_l = n_f-1$ quarks 
are considered as massless. 

As mentioned above, the $\overline{\rm MS}$ renormalized polarization
function is needed in Landau gauge. However, in our calculation we 
adopt a  general gauge parameter $\xi$ since the complexity is 
comparable to Landau gauge.

\begin{figure}[t]
  \begin{center}
    \begin{tabular}{c}
      \epsfig{file=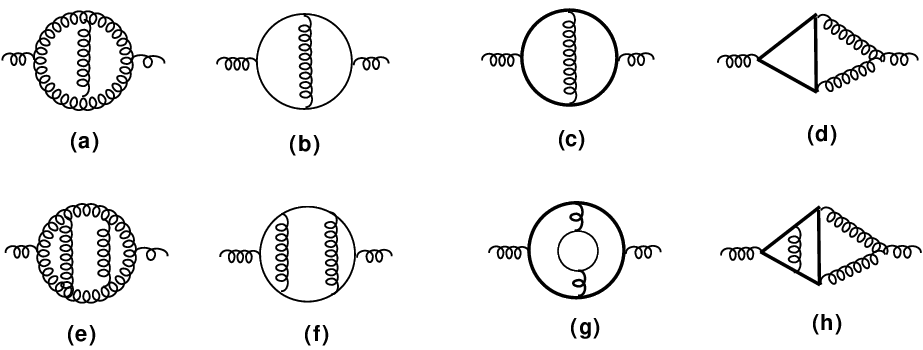,width=.95\textwidth}
    \end{tabular}
  \end{center}
%  \vspace{-4cm}
  \caption{\label{fig::bfm1}Sample diagrams contributing to the gluon
    propagator in the background field formalism at the two- and three-loop
    orders. 
    Diagrams (a), (b), (e) and (f) only contain massless lines, while the
    others also contain massive ones due to the presence of the heavy-quark
    loop (thick line). 
    (We have used the package 
    {\tt JaxoDraw}~\cite{Binosi:2003yf,Vermaseren:1994je} to draw the diagrams.) 
    }
\end{figure}

Some sample diagrams for $\Pi(q^2)$ are shown in Fig.~\ref{fig::bfm1}.
The diagrams are divided into two classes: completely massless diagrams
and 
and diagrams involving massive-quark loops.
The only scale in the massless diagrams is the external momentum. Thus they
can be evaluated using {\tt MINCER}~\cite{Gorishnii:1989gt,Larin:1991fz}.
In the second class, the mass of the heavy quark sets another scale which makes
the calculation significantly more difficult.
The one- and two-loop calculations can be performed analytically, and
the results can be found in Ref.~\cite{Jegerlehner:1998zg}.
At the three-loop order, however, an exact calculation is not yet possible.
We perform an asymptotic expansion in the limits $q^2\ll M_Q^2$ and 
$q^2\gg M_Q^2$. Note that due to the diagrams containing massive-quark loops
along with massless cuts (see, e.g., Figs.~\ref{fig::bfm1}(d) and (h))
also the small-$q^2$ expansion turns out to be nontrivial. As a result,
one encounters $\ln(q^2/M_Q^2)$ terms also in this limit.

All Feynman diagrams are generated with {\tt
  QGRAF}~\cite{Nogueira:1991ex}. The various
diagram topologies are identified and transformed to {\tt
  FORM}~\cite{Vermaseren:2000nd} with the help of {\tt q2e} and {\tt
  exp}~\cite{Harlander:1997zb,Seidensticker:1999bb}.  The program {\tt
  exp} is also used in order to apply the asymptotic expansion (see,
e.g., Ref.~\cite{Smirnov:2002pj}) in the various mass hierarchies. The
actual evaluation of the integrals is performed with the packages
{\tt MATAD}~\cite{Steinhauser:2000ry} and 
{\tt MINCER}~\cite{Larin:1991fz}, resulting in an expansion in
$d-4$ for each diagram, where $d$ is the space-time dimension.  

We computed four expansion terms for small and six
terms for large external momentum. In the following, we present only the
leading and subleading terms of the corresponding expansions for the gluon
polarization operator and the MOM $\beta$ function 
for $\mu^2=Q^2$ and provide the complete expressions in a
{\tt Mathematica} file.\footnote{See 
  {\tt
    http://www-ttp.particle.uni-karlsruhe.de/Progdata/ttp08/ttp08-50}.}
For completeness, we also list the one- and two-loop results, 
which agree with
the corresponding expansions of the exact
expressions~\cite{Jegerlehner:1998zg}.
It is convenient to cast the result in the form
\begin{eqnarray}
  \Pi(q^2) = 
  \Pi^{\rm ml}(q^2) 
  +  \Pi^{\rm mv}(q^2,M_Q^2)
  \,,
  \label{eq::Pi}
\end{eqnarray}
and introduce the variable
\begin{eqnarray}
  Z=\frac{Q^2}{4 M_Q^2}
  \,.
\end{eqnarray}

The results for the massless $\overline{\rm MS}$ renormalized polarization
function reads 
\begin{eqnarray}
  \Pi^{(1),{\rm ml}}_0 &=&
  -\frac{205}{48} + n_l\frac{5}{18}
  \,,
  \nonumber\\
  \Pi^{(2),{\rm ml}}_0 &=&
  -\frac{2687}{128} + \frac{513}{128} \zeta(3) 
  + n_l\left(\frac{347}{144} + \frac{1}{12}\zeta(3)\right)
  \,,
  \nonumber\\
  \Pi^{(3),{\rm ml}}_0 &=&
  - \frac{413343}{2048} 
  + \frac{58317}{1024}\zeta(3)
  + \frac{24885}{4096}\zeta(5)
  + n_l \left(\frac{1476013}{41472} - \frac{3797}{864}\zeta(3) 
  - \frac{145}{36}\zeta(5) \right)
  \nonumber\\&&\mbox{}
  + n_l^2\left(-\frac{64627}{62208} - \frac{1}{432}\zeta(3)\right)
  \,.
\end{eqnarray}
In the limit $Z \to 0$, we get 
\begin{eqnarray}
  \Pi^{(1),{\rm mv}}_0 &=&
  \frac{1}{6}\lnfourZ - \frac{2}{15}Z
  +{\cal O}(Z^2)
  \,,
  \nonumber\\
  \Pi^{(2),{\rm mv}}_0 &=&
  \frac{7}{24} + \frac{19}{24}\lnfourZ 
  + \left(\frac{50239}{97200} - \frac{7}{15}\lnfourZ\right)Z
  +{\cal O}(Z^2)
  \,,
  \nonumber\\
  \Pi^{(3),{\rm mv}}_0 &=&
  \frac{58933}{124416} 
  + \left(\frac{2}{3}+\frac{2}{9}\ln2\right)\zeta(2)
  + \frac{80507}{27648}\zeta(3)
  + \lnfourZ\left(\frac{58939}{ 6912} 
  - \frac{171}{256}\zeta(3)
  \right)
  \nonumber\\&&\mbox{}
  + \frac{283}{576}\lnfourZtwo
  + n_l\left[-\frac{2479}{31104} - \frac{1}{9}\zeta(2) 
  + \lnfourZ\left(-\frac{1103}{1728} - \frac{1}{72}\zeta(3)\right)
  \right]
  \nonumber\\
%%%%%%%%%%%%%%%%%%%%%%%%%%%%%
%&{+}&
&&\mbox{}+
Z\,\left\{
\frac{6252381359}{279936000} 
+\frac{495461}{2332800} \,\ln(4Z) 
-\frac{6403}{4608}  \, \ln^2(4Z)
-\frac{8}{15}  \,\zeta(2)
\right.
\nonumber\\
&{}&
\left.
%\rule{1cm}{0mm}
-\frac{8}{45} \,\zeta(2)\,\ln2  
-\frac{625415}{31104}  \,\zeta(3)
+n_l 
\left[
-\frac{118427}{291600} 
+\frac{12401}{58320} \,\ln(4Z) 
%zero == 0
\right.
\right.
\nonumber\\
%&{+}&\,
&&\mbox{}
\left.
\left.
+\frac{61}{2592}  \, \ln^2(4Z)
+\frac{4}{45}  \,\zeta(2)
%zero == 0
\right]
\right\}
%%%%%%%%%%%%%%%%%%%%%%%%%%%%
  +{\cal O}(Z^2)
\label{smallZ} 
 \,,
\end{eqnarray}
\ice{ (* 

In[108]:=  Coll[Coef[ressmaOSh,as,3]//cut[Z,{0,0}],{nl,log4Z}]

                            2
          58933    283 log4Z    2 z2   2 log2 z2
Out[108]= ------ + ---------- + ---- + --------- + 
          124416      576        3         9
 
           2479     z2            1103    z3            58939   171 z3
>    nl (-(-----) - -- + log4Z (-(----) - --)) + log4Z (----- - ------) + 
           31104    9             1728    72            6912     256
 
     80507 z3
>    --------
      27648

 *)
}
and in the large-$Z$ region, we obtain
\begin{eqnarray}
  \Pi^{(1),{\rm mv}}_0 &=&
  \frac{5}{18} - \frac{1}{4Z}
  + {\cal O}\left(\frac{1}{Z^2}\right)
  \,,
  \nonumber\\
  \Pi^{(2),{\rm mv}}_0 &=&
  \frac{347}{144} + \frac{1}{12}\zeta(3) 
  - \frac{1}{Z}\left(
  \frac{233}{128} 
  + \frac{3}{8}\zeta(3)
  - \frac{5}{64}\lnfourZ 
  \right) 
  + {\cal O}\left(\frac{1}{Z^2}\right)
  \,,
  \nonumber\\
  \Pi^{(3),{\rm mv}}_0 &=&
  \frac{4298785}{124416} 
  - \frac{3799}{864}\zeta(3)
  - \frac{145}{36}\zeta(5)
  + n_l\left(-\frac{64627}{31104} - \frac{1}{216}\zeta(3)\right)
   \nonumber\\
%&{+}&
&&\mbox{}+
\frac{1}{Z}
\left\{
-\frac{187715}{6912} 
+\frac{79355}{18432} \,\ln(4Z) 
-\frac{3127}{6144}  \, \ln^2(4Z)
+  \,\zeta(2)
+\frac{1}{3} \,\zeta(2)\,\ln2  
\right. \nonumber \\ &&
%\BreakI
%\phantom{+}
-\frac{349853}{27648}  \,\zeta(3)
+\frac{15}{128} \,\zeta(3) \,\ln(4Z)  
+\frac{81}{64}  \,\zeta(4)
+\frac{23425}{3456}  \,\zeta(5)
%zero == 0
+n_l 
\left[
\frac{785}{576} 
\right.
\nonumber\\
%&{+}&\,
&&\mbox{}
\left.\left.
-\frac{181}{1152} \,\ln(4Z) 
+\frac{11}{384}  \, \ln^2(4Z)
-\frac{1}{6}  \,\zeta(2)
+\frac{89}{96}  \,\zeta(3)
%zero == 0
\right]
\right\}
+ {\cal O}\left(\frac{1}{Z^2}\right)
\,.
\label{largeZ} 
\end{eqnarray}
\ice{
In[124]:= Coll[Coef[reslarOSh/.Z->1/iZ,as,3]//cut[iZ,{0,0}],{nl,log4Z,iZ}]

          4298785         64627    z3     3799 z3   145 z5
Out[124]= ------- + nl (-(-----) - ---) - ------- - ------
          124416          31104    216      864       36
}

The small- and large-$Z$ expansions   of the MOM $\beta$ function read:
\bea
\betaMOM_0 &\bbuildrel{=\!=\!=}_{Z \to 0}^{} & 
\frac{11}{4}  -\frac{n_l}{6} -Z\frac{2}{15} 
+ {\cal O}\left(Z^2\right)
\nonumber
{},
\\
\betaMOM_1 &\bbuildrel{=\!=\!=}_{Z \to 0}^{}&
\frac{51}{8}- n_l\,\frac{19}{24}
+ Z\left(
\frac{4879}{97200}-\frac{7}{15}\ln(4Z)
\right)
+ {\cal O}\left({Z^2}\right)
\nonumber
{},
\\
%\eea
%\bea
\betaMOM_2 &\bbuildrel{=\!=\!=}_{Z \to 0}^{}&
\frac{27045}{512} 
-\frac{5643}{512}  \,\zeta(3)
%zero == 0
{-}\,n_l 
\left[
\frac{7175}{768} 
-\frac{337}{768}  \,\zeta(3)
%zero == 0
\right]
{+} \, n_l^2
\left[
\frac{953}{3456} 
+\frac{1}{72}  \,\zeta(3)
%zero == 0
\right]
\nonumber\\
&&\mbox{}+
Z\, \left\{
\frac{5640101219}{279936000} 
+\frac{995773}{1555200} \,\ln(4Z) 
-\frac{11269}{7680}  \, \ln^2(4Z)
-\frac{8}{15}  \,\zeta(2)
\right. \nonumber \\ && \mbox{}
%\rule{4mm}{0mm}
%\phantom{+}
-\frac{8}{45} \,\zeta(2)\,\ln2  
-\frac{625415}{31104}  \,\zeta(3)
%zero == 0
+n_l 
\left[
\frac{1063}{87480} 
+\frac{305}{5832} \,\ln(4Z) 
\right.
\nonumber\\
&&\mbox{}
\left.\left.
+\frac{61}{2592}  \, \ln^2(4Z)
+\frac{4}{45}  \,\zeta(2)
%zero == 0
\right]
\right\}
+ {\cal O}\left({Z^2}\right)
{}\,,
\label{betaMOM2sma}
\eea

\bea
\betaMOM_0 &\bbuildrel{=\!=\!=}_{Z \to \infty}^{} & 
\frac{31}{12}  -\frac{n_l}{6} +\frac{1}{4 Z} 
+ {\cal O}\left(\frac{1}{Z^2}\right)
{},
\nonumber
\\
\betaMOM_1 &\bbuildrel{=\!=\!=}_{Z \to \infty}^{}&
\frac{67}{12}- n_l\,\frac{19}{24}
+ \frac{1}{Z}\left(
\frac{243}{128}-\frac{5}{64}\ln(4Z)
+\frac{3}{8} \zeta(3)
\right)
+ {\cal O}\left(\frac{1}{Z^2}\right)
\nonumber
{},
\\
%\eea
%\bea
\betaMOM_2 &\bbuildrel{=\!=\!=}_{Z \to \infty}^{} & 
\frac{604877}{13824} 
-\frac{48701}{4608}  \,\zeta(3)
%zero == 0
{+}\,n_l 
\left[
-\frac{60763}{6912} 
+\frac{1075}{2304}  \,\zeta(3)
%zero == 0
\right]
{+} \, n_l^2
\left[
\frac{953}{3456} 
+\frac{1}{72}  \,\zeta(3)
%zero == 0
\right]
\nonumber\\
&&\mbox{}+
\frac{1}{Z}\, \left\{
\frac{751727}{27648} 
-\frac{32029}{6144} \,\ln(4Z) 
+\frac{3127}{6144}  \, \ln^2(4Z)
-  \,\zeta(2)
-\frac{1}{3} \,\zeta(2)\,\ln2  
\right. \nonumber \\ &{}& \mbox{}
+\frac{338477}{27648}  \,\zeta(3)
-\frac{15}{128} \,\zeta(3)\,\ln(4Z)  
-\frac{81}{64}  \,\zeta(4) 
-\frac{23425}{3456}  \,\zeta(5)        
%zero == 0
+ \,n_l 
\left[
-\frac{1265}{1152} 
\right.
\nonumber\\
&&\mbox{}
\left.
\left.
+\frac{79}{384} \,\ln(4Z) 
-\frac{11}{384}  \, \ln^2(4Z)
+\frac{1}{6}  \,\zeta(2)
-\frac{85}{96}  \,\zeta(3)
%zero == 0
\right]
\right\}
+ {\cal O}\left(\frac{1}{Z^2}\right)
{}.
\label{betaMOM2lar}
\eea

Note that our result for the MOM $\beta$ function explicitly demonstrates 
the validity of the Applelquist-Carazonne theorem \cite{Appelquist:1974tg} 
at the three-loop level. Indeed, one can easily check that, for $i=0,1$ 
and 2, one has
\bea
\lim_{Z \to 0} \betaMOM_i  &\equiv& \betaMOM_{i,{\rm ml}}(n_f = n_l)
{},
\nonumber
\\
\lim_{Z \to \infty } \betaMOM_i  &\equiv& \betaMOM_{i,{\rm ml}}(n_f = n_l+1)
\nonumber
{},
\eea
where $\betaMOM_{i,{\rm ml}}(n_f)$ is the three-loop contribution to the MOM
$\beta$ function in the massless limit (see Eq.~(\ref{eq::beta012})).

\begin{figure}[t]
  \begin{center}
    \begin{tabular}{c}
      \epsfig{file=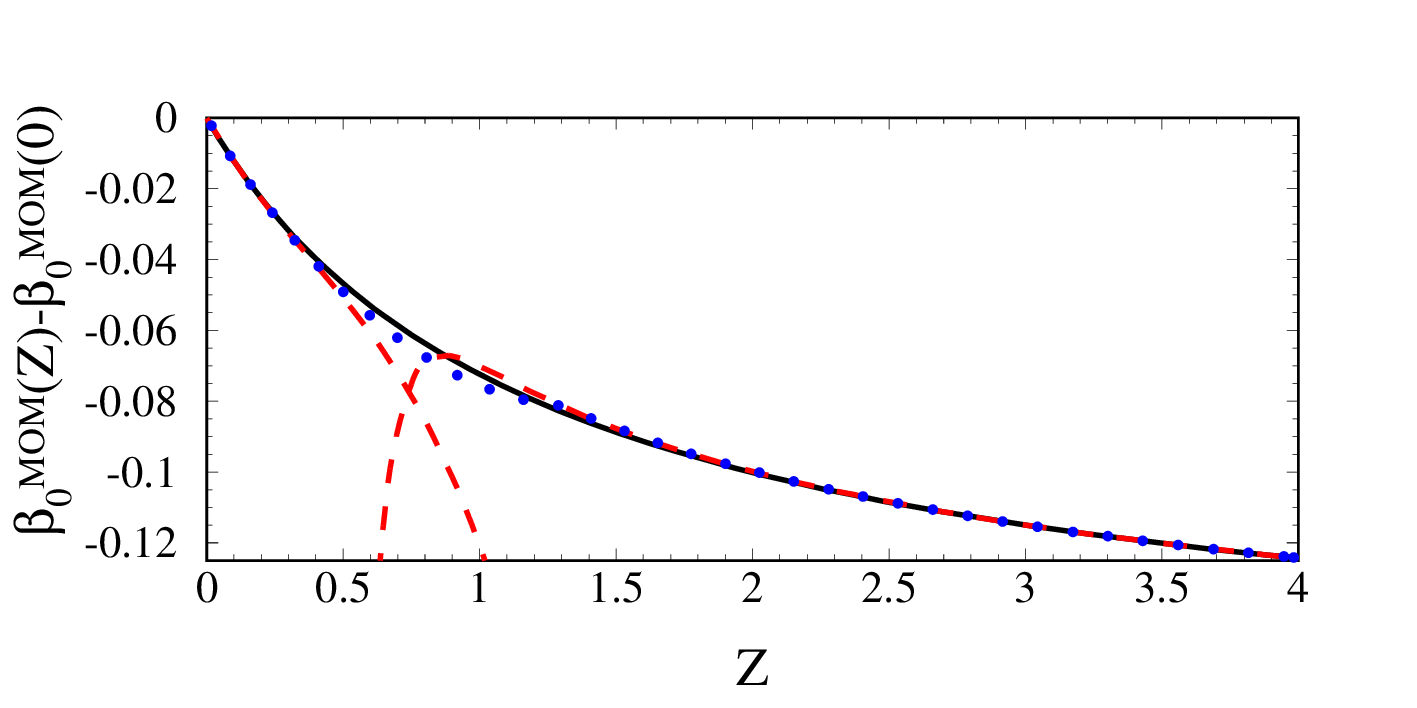,width=.78\textwidth}
      \\
      \vspace*{-3em}
      \\
      \epsfig{file=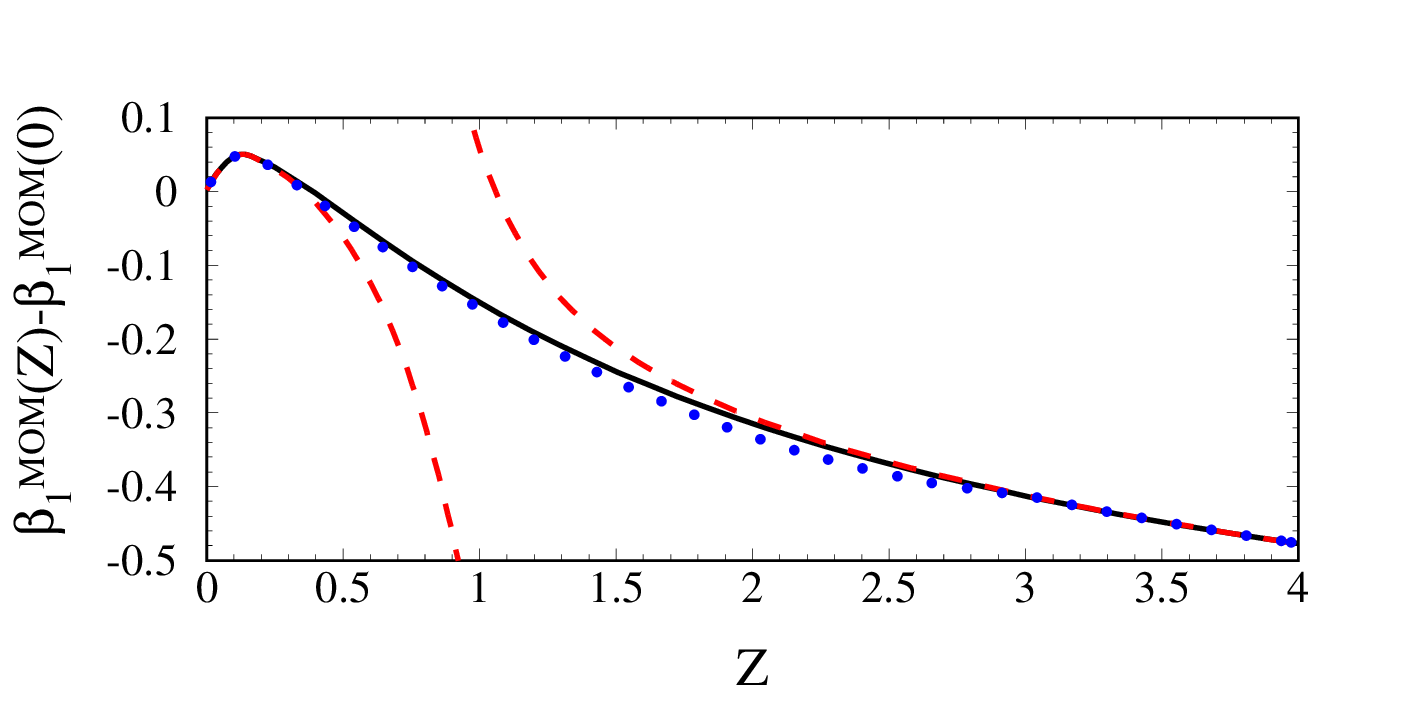,width=.78\textwidth}
      \\
      \vspace*{-3em}
      \\
      \epsfig{file=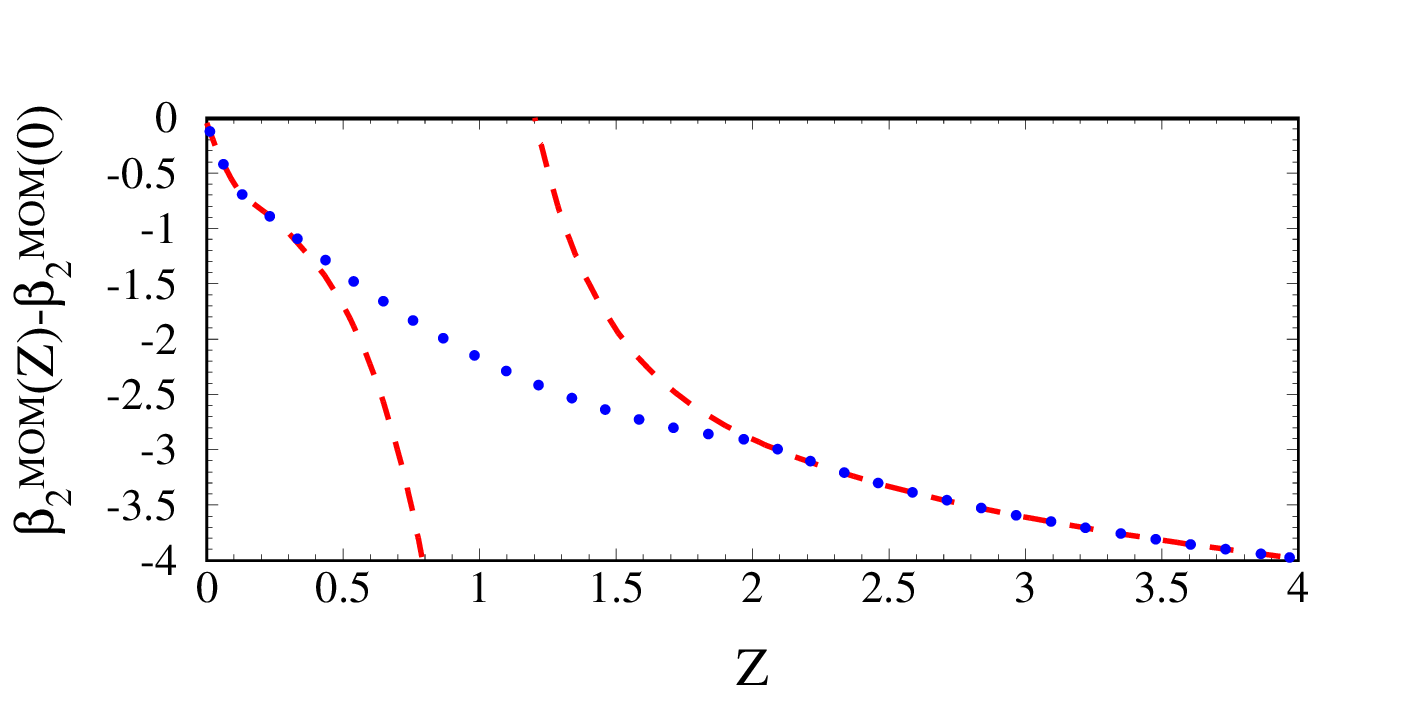,width=.78\textwidth}
    \end{tabular}
    \caption{\label{fig::beta} The difference 
      $\beta_i^{\rm MOM}(Z) - \beta_i^{\rm MOM}(0)$ ($i=0,1,2$) as a function of 
      $Z$ in the Euclidian region. 
      In the three-loop result, $n_l=n_f-1=5$ has been chosen. In each frame, 
      the dashed lines represent the approximation for small and large
      values of $Z$, the dotted curve is the result of the
      interpolation, and the solid line (for $\beta_0^{\rm MOM}$ and
      $\beta_1^{\rm MOM}$) is the exact result from Ref.~\cite{Jegerlehner:1998zg}.
   }
  \end{center}
\end{figure}

In Fig.~\ref{fig::beta}, we present the results for the MOM $\beta$
function in graphical form, where the one-, 
two- and three-loop coefficients are shown as functions of 
$Z$ in the Euclidian region. Next to the low- and high-energy
approximations (dashes) including the $Z^3$ and $1/Z^5$ terms,
also the interpolation functions (dotted) are shown.
At the one- and two-loop orders, these results are compared against the
exact result (solid line).
For demonstration purpose, we have chosen $n_l=5$ at the three-loop order.
Very similar results are obtained for other values of $n_l$.

%- }}}
%- {{{ Phenomenological applications:

\section{\label{sec::applications}Phenomenological applications}

In the following, we discuss the numerical impact of the results obtained in
this paper. In particular, we consider the $\overline{\rm MS}$ quantity
$\alpha_s^{(5)}(M_Z)$ as input value and evaluate the strong coupling at lower
and higher energy scales with different numbers of active flavours. On the one
hand, this can be done in the $\overline{\rm MS}$ scheme applying the usual
running and decoupling procedure (see, e.g.,
Refs.~\cite{Chetyrkin:1997un,Steinhauser:2002rq}). In this case, one has to
specify a scale $\mu_Q$ where the heavy quark $Q$ is integrated out.
On the other hand, it is possible to switch from the $\overline{\rm MS}$ to the
MOM ($\overline{\rm MOM}$) scheme for $\mu=M_Z$ and perform the running with
the help of the MOM ($\overline{\rm MOM}$) $\beta$ function.
The results obtained at lower and higher energies can also
be translated back to the $\overline{\rm MS}$ scheme, and a comparison can be
performed. In this way, we can check the consistency 
between the two renormalization schemes.

\begin{table}[t]
\begin{center}
\begin{tabular}{r|r|r|r|r|r|r|r|r}
  \multicolumn{1}{c|}{$\mu$~(GeV)} & $n_f$ &
  \multicolumn{2}{c|}{$\alpha_s^{(n_f)}(\mu)$} &
  \multicolumn{1}{c|}{$n_f^{\rm MOM}$} &
  \multicolumn{2}{c|}{$\alphaMOM_s(\mu)$} &
  \multicolumn{2}{c}{$\alsMOM(\alpha_s(\mu))$} \\
 \hline 
  & & 2 loop & 3 loop &
  &
  2 loop & 3 loop & 
  2 loop & 3 loop 
  \\ 
\hline
     91.19& 5 &  0.1180&  0.1180& 6&  0.1324&  0.1331&  0.1324&0.1331 \\
      200& 5 &  0.1055&  0.1055& 6&  0.1170&  0.1175&  0.1171&      0.1175 \\
      350& 6 &  0.0989&  0.0990& 6&  0.1082&  0.1086&  0.1084&      0.1087 \\
      500& 6 &  0.0950&  0.0951& 6&  0.1034&  0.1037&  0.1036&      0.1038 \\
      1000& 6 &  0.0884&  0.0885& 6&  0.0953&  0.0956&  0.0955&      0.0957 \\
 \end{tabular}
\caption{\label{A:MOM}$\alpha_s$ and $\alphaMOM_s$  for various
  values of $\mu$ from region~A. As input, $\alpha_s(M_Z)=0.118$ is used,
  which is transformed with the help of Eq.~(\ref{eq::asMOMvsMS}) to
  $\alphaMOM_s(M_Z)$. The running to different values of $\mu$ is
  achieved with the help of the appropriate $\beta$ function.
  $\alsMOM(\alpha_s(\mu))$ is obtained from $\alpha_s(\mu)$ using
  Eq.~(\ref{eq::asMOMvsMS}).}
\end{center}
%\end{table}
%\begin{table}[t]
\begin{center}
\begin{tabular}{r|r|r|r|r|r|r|r|r}
  \multicolumn{1}{c|}{$\mu$~(GeV)} & $n_f$ &
  \multicolumn{2}{c|}{$\alpha_s^{(n_f)}(\mu)$} &
  \multicolumn{1}{c|}{$n_f^{\rm MOM}$} &
  \multicolumn{2}{c|}{$\alsMOMbar(\mu)$} &
  \multicolumn{2}{c}{$\alsMOMbar(\alpha_s(\mu))$} \\
 \hline 
  & & 2 loop & 3 loop &
  &
  2 loop & 3 loop & 
  2 loop & 3 loop 
  \\
\hline
    91.19& 5 &  0.1180&  0.1180& 6&  0.1192&  0.1196&  0.1192&0.1196 \\
     200& 5 &  0.1055&  0.1055& 6&  0.1066&  0.1068&  0.1066&  0.1068     \\
     350& 6 &  0.0989&  0.0990& 6&  0.0993&  0.0995&  0.0993&  0.0995     \\
     500& 6 &  0.0950&  0.0951& 6&  0.0952&  0.0954&  0.0952&  0.0954     \\
     1000& 6 &  0.0884&  0.0885& 6&  0.0884&  0.0885&  0.0884&     0.0885 \\
 \end{tabular}
\caption{\label{A:MOMbar}$\alpha_s$ and $\alsMOMbar$ for various
  values of $\mu$ from region~A. As input, $\alpha_s(M_Z)=0.118$ is used,
  which is transformed with the help of Eq.~(\ref{eq::asMOMBvsMS})
  to $\alsMOMbar(M_Z)$. The running to different values of $\mu$ is
  achieved with the help of the appropriate $\beta$ function.
  $\alsMOMbar(\alpha_s(\mu))$ is obtained from $\alpha_s(\mu)$ using
  Eq.~(\ref{eq::asMOMBvsMS}).}
\end{center}
\end{table}

\begin{table}[t]
\begin{center}
\begin{tabular}{r|r|r|r|r|r|r|r|r}
  \multicolumn{1}{c|}{$\mu$~(GeV)} & $n_f$ &
  \multicolumn{2}{c|}{$\alpha_s^{(n_f)}(\mu)$} &
  \multicolumn{1}{c|}{$n_f^{\rm MOM}$} &
  \multicolumn{2}{c|}{$\alphaMOM_s(\mu)$} &
  \multicolumn{2}{c}{$\alsMOM(\alpha_s(\mu))$} \\
 \hline 
  & & 2 loop & 3 loop &
  &
  2 loop & 3 loop & 
  2 loop & 3 loop 
  \\ 
\hline
     91.19& 5 &  0.1180&  0.1180& 5&  0.1328&  0.1332&  0.1328&0.1332 \\
      50& 5 &  0.1298&  0.1298& 5&  0.1480&  0.1486&  0.1480&  0.1486      \\
      10& 5 &  0.1779&  0.1781& 5&  0.2177&  0.2198&  0.2160&  0.2192      \\
      4& 4 &  0.2288&  0.2287& 5&  0.3074&  0.3148&  0.2988&  0.3096      \\
      3& 4 &  0.2536&  0.2538& 5&  0.3556&  0.3682&  0.3404&  0.3574      \\
 \end{tabular}
\caption{\label{B:MOM}$\alpha_s$ and $\alphaMOM_s$ for various
  values of $\mu$ from region~B. As input, $\alpha_s(M_Z)=0.118$ is used,
  which is transformed with the help of Eq.~(\ref{eq::asMOMvsMS})
  to $\alphaMOM_s(M_Z)$. The running to different values of $\mu$ is
  achieved with the help of the appropriate $\beta$ function.
  $\alsMOM(\alpha_s(\mu))$ is obtained from $\alpha_s(\mu)$ using 
  Eq.~(\ref{eq::asMOMvsMS}).}
\end{center}
%\end{table}
%\begin{table}[t]
\begin{center}
\begin{tabular}{r|r|r|r|r|r|r|r|r}
  \multicolumn{1}{c|}{$\mu$~(GeV)} & $n_f$ &
  \multicolumn{2}{c|}{$\alpha_s^{(n_f)}(\mu)$} &
  \multicolumn{1}{c|}{$n_f^{\rm MOM}$} &
  \multicolumn{2}{c|}{$\alsMOMbar(\mu)$} &
  \multicolumn{2}{c}{$\alsMOMbar(\alpha_s(\mu))$} \\
 \hline 
  & & 2 loop & 3 loop &
  &
  2 loop & 3 loop & 
  2 loop & 3 loop 
  \\
\hline
    91.19& 5 &  0.1180&  0.1180& 5&  0.1180&  0.1181&  0.1180&0.1181 \\
     50& 5 &  0.1298&  0.1298& 5&  0.1298&  0.1300&  0.1298&  0.1300     \\
     10& 5 &  0.1779&  0.1781& 5&  0.1797&  0.1804&  0.1798&  0.1807     \\
     4& 4 &  0.2288&  0.2287& 5&  0.2350&  0.2376&  0.2352&  0.2385     \\
     3& 4 &  0.2536&  0.2538& 5&  0.2612&  0.2652&  0.2615&  0.2667     \\
 \end{tabular}
\caption{\label{B:MOMbar}$\alpha_s$ and $\alsMOMbar$ for various
  values of $\mu$ from region~B. As input, $\alpha_s(M_Z)=0.118$ is used,
  which is transformed with the help of Eq.~(\ref{eq::asMOMBvsMS})
  to $\alsMOMbar(M_Z)$. The running to different values of $\mu$ is
  achieved with the help of the appropriate $\beta$ function.
  $\alsMOMbar(\alpha_s(\mu))$ is obtained from $\alpha_s(\mu)$ using
  Eq.~(\ref{eq::asMOMBvsMS}).}
\end{center} 
\end{table}

For our numerical analysis, we use the following input values
\begin{eqnarray}
  \alpha_s^{(5)}(M_Z)=0.118\,,\quad
  M_b=4.7~\mbox{GeV}\,,\quad 
  M_t=175~\mbox{GeV}\,,
\end{eqnarray}
where $M_Q$ represent the pole quark masses and
for the decoupling scales we choose $\mu_Q= 2\,M_Q$.
The running and decoupling in the $\overline{\rm MS}$ scheme is performed with
the help of {\tt RunDec}~\cite{Chetyrkin:2000yt}.

We consider two regions of energies. Region~A starts from the $Z$-boson mass, $M_Z$,
and extends to energies much higher than the top-quark mass, say, 1000 GeV.
In this region, we investigate the evolution of the strong-coupling constant 
in the MOM and $\overline{\rm MOM}$ schemes with five massless and one
heavy quark, the top quark. Thus in total six quarks are present
in the theory which we denote as $n_f^{\rm MOM}=6$.
On the other hand, in region~B, we consider the evolution of $\alphaMOM_s(\mu)$
and $\alsMOMbar(\mu)$ from $\mu=M_Z$  down to $\mu = 3$ GeV. The number of
massless quarks for region B is set to four, and the heavy quark be should 
identified with the bottom quark, i.e., we have $n_f^{\rm MOM}=5$.

In Tab.~\ref{A:MOM}, we compare the values for $\alpha_s$ for some selected
$\mu$ values from
region~A in the $\overline{\rm MS}$ and MOM
schemes. For all numbers, we choose $\alpha_s^{(5)}(M_Z)$ as the starting
point, transform at $\mu=M_Z$ to the {\rm MOM}
scheme and use the corresponding renormalization group equation to
arrive at the desired $\mu$ values.
The same comparison for the case of the $\overline{\rm MOM}$ scheme is shown 
Tab.~\ref{A:MOMbar}. 
In both tables, we show in the last two columns the results of
$\alphaMOM_s(\mu)$ (Tab.~\ref{A:MOM}) and $\alsMOMbar(\mu)$
(Tab.~\ref{A:MOMbar}) as obtained from the $\overline{\rm MS}$-evolved value 
$\alpha_s(\mu)$ using
Eqs.~(\ref{eq::asMOMvsMS}) and~(\ref{eq::asMOMBvsMS}), respectively.
The corresponding results for region~B are shown in Tabs.~\ref{B:MOM}
and~\ref{B:MOMbar} (where, of course, the values given in
Eq.~(\ref{eq::nfMOM5values}) have been used).

All four tables show good agreement between the values of the MOM coupling
constant obtained via direct integration of the (quark-mass-dependent) MOM
$\beta$ function and with the help of the (simpler) conversion from
the $\msbar$ scheme.  
For Tabs.~\ref{A:MOM} and~\ref{B:MOM} the agreement is even getting better
after taking into account the three-loop corrections.
Note that the results in the case of the $\overline{\rm MOM}$ scheme
become slightly worse after switching on the three-loop terms,
as can be seen in Tabs.~\ref{A:MOMbar} and~\ref{B:MOMbar}.
The reason for this can be seen by comparing
Eq.~(\ref{asMOM_through_asMS:nl6,nl,rescaled}) with
Eq.~(\ref{asMOM_through_asMS:nl6,nl}). The former has (by
construction) vanishing order $\alpha_s$ corrections and a two-loop
coefficient which is smaller by a factor of six. However, the three-loop term is
only reduced by a factor of three and thus has bigger relative influence.
Still, the difference between $\alsMOMbar(\mu)$ and 
$\alsMOMbar(\alpha_s(\mu))$ for $\mu=3$~GeV is about a factor of ten less
than the current best value obtained, e.g., from hadronic $\tau$ decay
(see, e.g., Ref.~\cite{Baikov:2008jh}).

\begin{figure}[t]
  \begin{center}
    \epsfig{file=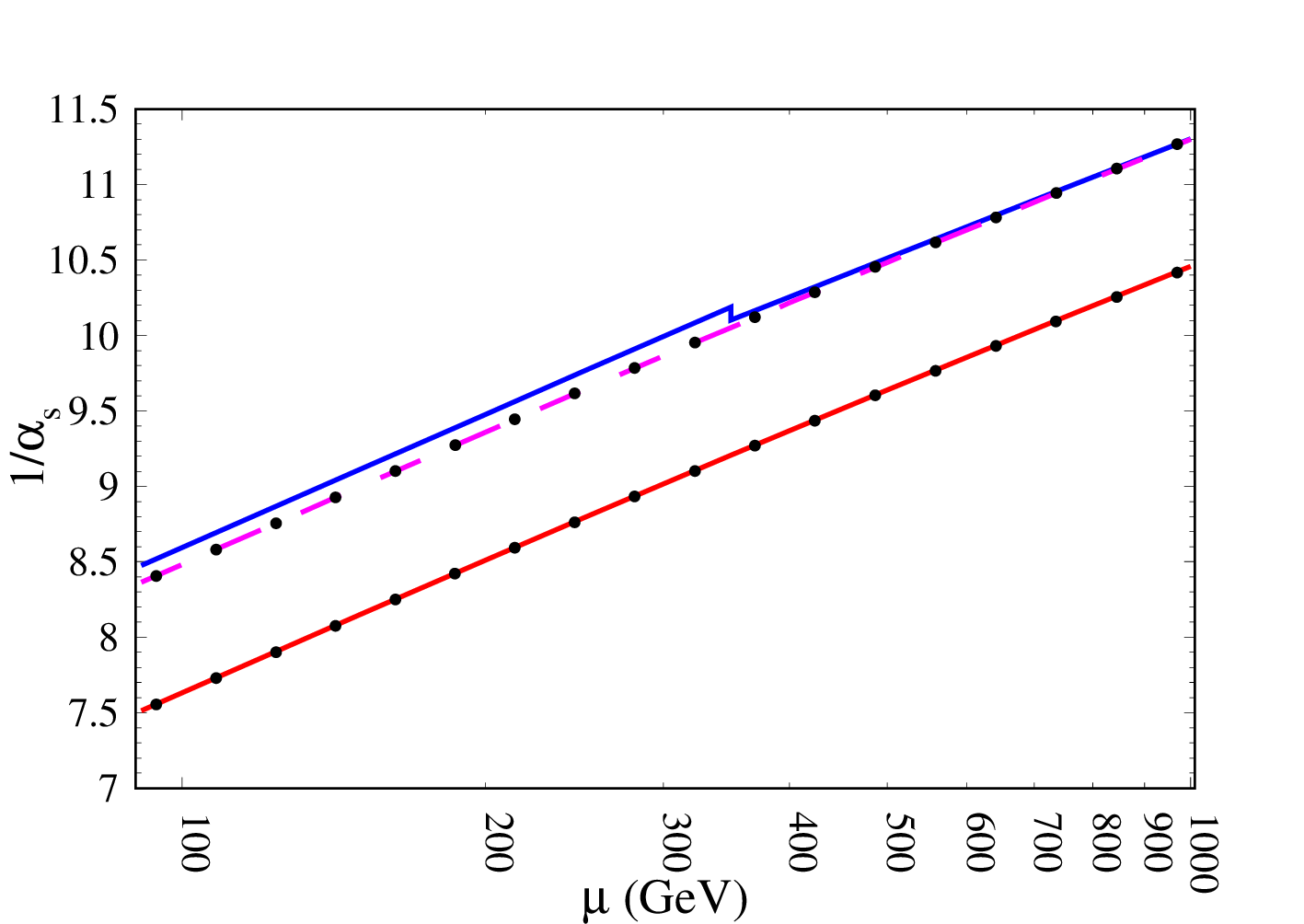,width=\textwidth}
  \end{center}
  \vspace*{-2em}
  \caption{\label{fig::asmu6}$1/\alpha_s$ as a function of $\mu$.
    The (blue) upper solid line containing a step for $\mu=2M_t$ corresponds to the
    $\overline{\rm MS}$ result and the (red) lower solid line to
    the result in the MOM scheme. The (pink) dashed line represents
    $1/\alpha_s$ in the $\overline{\rm MOM}$ scheme. 
    The (black) dotted lines lying on top of the MOM and $\overline{\rm MOM}$ result
    correspond to the results obtained from $\overline{\rm MS}$ value of
    $\alpha_s$ using the conversion formulae~(\ref{eq::asMOMvsMS})
    and~(\ref{eq::asMOMBvsMS}), respectively.
    For all results the three-loop expressions have been used and $n_f^{\rm
      MOM}=6$ has been chosen.
  }
\end{figure}

\begin{figure}[t]
  \begin{center}
    \epsfig{file=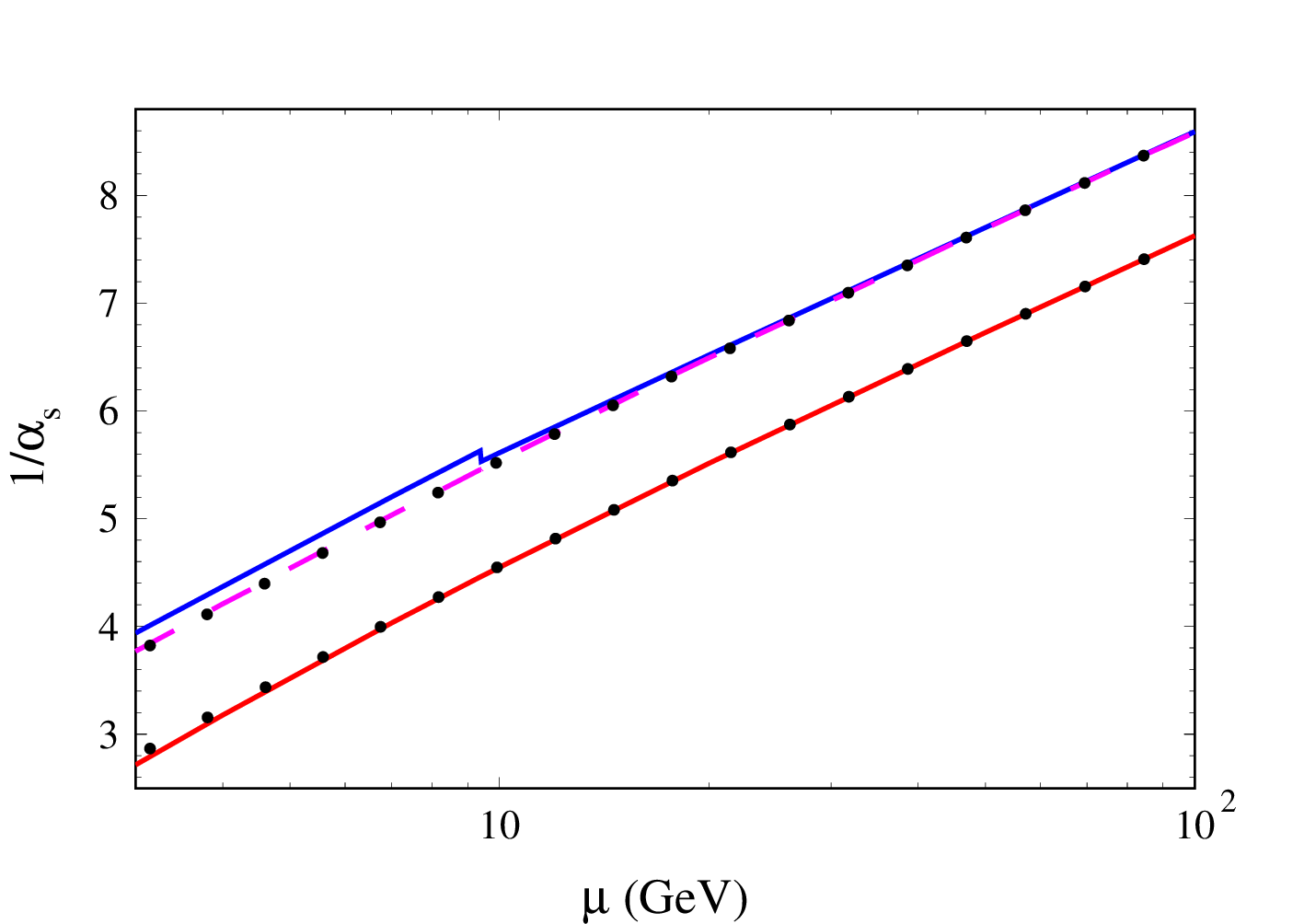,width=\textwidth}
  \end{center}
  \vspace*{-2em}
  \caption{\label{fig::asmu5} The same coding as in Fig.~\ref{fig::asmu6}
    has been adapted, except that $n_f^{\rm MOM}=5$ has been chosen.
  }
\end{figure}

In Figs.~\ref{fig::asmu6} and~\ref{fig::asmu5}, the results of
Tabs.~\ref{A:MOM}--\ref{B:MOMbar} are shown in graphical from. In particular, 
we plot the inverse strong coupling as a
function of $\mu$ both for the $\overline{\rm MS}$, MOM and 
$\overline{\rm MOM}$ schemes, where in all
cases the three-loop approximation is used for the running and the conversion
between the schemes. 
We again choose $\alpha_s^{(5)}(M_Z)$ as the input quantity and convert at this
scale to the other two schemes.
The evolution of the $\overline{\rm MS}$ coupling to
lower $\mu$ values is shown by the (upper) solid lines with a step at the
values for $\mu_t=2M_t$ and $\mu_b=2M_b$, respectively. 
Numerically very close is the dashed curve in the $\overline{\rm MOM}$ scheme, 
which is expected from the above discussion. The lower solid line represents the
result in the MOM scheme. Both for the {\rm MOM} and $\overline{\rm
  MOM}$ results, the conversion is performed for $\mu=M_Z$, and the
running to other values of $\mu$ is achieved using the corresponding
$\beta$ function. The dotted lines on top of the 
MOM and $\overline{\rm MOM}$ curves represent the results where the
transformation from the $\overline{\rm MS}$ values is performed just at the
considered value of $\mu$. 

%- }}}
%- {{{ Conclusions

\section{\label{sec::concl}Conclusions}

We have computed the three-loop corrections to the $\beta$
function of QCD with one heavy and $n_l$ massless quarks
in a momentum subtraction scheme (MOM). In
our three-loop calculation, we do not consider the diagrams involving 
two different quark masses. Although there are only a few diagrams of
this type, their evaluation is significantly more difficult.

We have shown that our results describe the MOM coupling constant evolution in
well-defined kinematical regions with three-loop accuracy. 
Moreover, the numerical analysis of our results has clearly
demonstrated the full equivalence of the schemes with explicitly built-in
decoupling (MOM and $\overline{\mbox{MOM}}$) to the standard $\msbar$ scheme,
which, as is well-known, does not have such a property.

From the more technical point of view, it has been shown that one can use the
$\msbar$ scheme evolution along with simple conversion relations
(derived for the regions either significantly above or below the heavy-quark
threshold) to relate the values of the MOM scheme coupling constant  
from both regions. 

Finally, we believe that our analysis should help to remove the last
traces of doubt about the usefulness of $\msbar$-like schemes, which formally
do not obey the the Applelquist-Carazzone theorem~\cite{Appelquist:1974tg}, 
for a unified description of mass effects in a broad region of $\mu^2$ values,
from far below heavy-quark thresholds to well beyond them.

%- }}}
%- {{{ Acknowledgments:

\bigskip
\noindent
{\large\bf Acknowledgments}\\
This work was supported in part by the BMBF through Grant Nos. 05\,HT6VKA
and 05\,HT6GUA and by the DFG through SFB/TR~9.

%- }}}
%- {{{ Bibliography

%- }}}

\end{document}